\documentclass[twocolumn,secnumarabic,amssymb, nobibnotes, aps, prd]{revtex4-1}
\usepackage{ulem}
\usepackage{color}
\usepackage[colorlinks]{hyperref}
\usepackage{amssymb}
\usepackage{amsmath}
\usepackage{animate}
\usepackage{graphicx}
\usepackage{mathrsfs}
\usepackage{bm}
\usepackage{picinpar}

\begin{document}

\title[Influence of a dark-soliton on the reflection of a BEC by a square barrier]{Influence of a dark-soliton on the reflection of a Bose-Einstein condensate by a square barrier}%
\author{Qiao-Ling Cheng$^{1}$, Wen-Kai Bai$^{1}$, Yao-Zhong Zhang$^{2}$, Bo Xiong$^{3}$ and Tao Yang$^{1,4}$}

\address{$^1$Institute of Modern Physics, Northwest University, Xi'an 710069, P. R. China}
\address{$^2$School of Mathematics and Physics, The University of Queensland, Brisbane Qld 4072, Australia}
\address{$^3$School of Science, Wuhan University of Technology, Wuhan 430070, P. R. China }
\email{boxiongpd@gmail.com}
\address{$^4$Shaanxi Key Laboratory for Theoretical Physics Frontiers, Xi¡¯an 710069, P. R. China}
\email{yangt@nwu.edu.cn}

\begin{abstract}
We study the quantum reflection of a two-dimensional disk-shaped Bose-Einstein condensate with a dark-soliton excitation by a square potential barrier. For the giving geometry, the dark-soliton initially located at the centre of the condensate cloud survive long enough for investigating the reflection process. We show the time evolution of the reflection probability with respect to various width of the barrier. The asymptotic value of the reflection probability is decreased by the existence of a dark-soliton, and is highly sensitive to the initial orientation of the dark-soliton which also affects the excitation properties during the process of condensate and barrier interaction.
\end{abstract}
\vspace{2pc}


\maketitle


\section{Introduction}

Solitons are, strictly speaking, shape-preserving excitations supported in nonlinear media, where the effect of dispersion is balanced by the interparticle interactions \cite{JPB.36.2891}. Dark-soliton appears as a local notch in the atomic density with a phase slip across it, whose effect is compensated by the quantum pressure term arising from the kinetic energy \cite{BEC-P}. It is directly related to the nonlinearity of the Gross-Pitaevskii (GP) equations. Generally this localized nonlinear wave can propagate over long distances without change in shape and has appeared in various systems \cite{JPA.43.213001}, such as optical fibers \cite{PRL.60.29,PRL.61.2445}, magnetic films \cite{PRL.70.1707}, plasmas \cite{PRL.102.135002}, and waveguide arrays \cite{pre.74.065601}. In the past decades, after the realization of atomic Bose-Einstein condensates (BECs), the dark-solitons in presence of matter-wave systems attracted much attention. In a condensate cloud, such solitons corresponding to a phase jump can be achieved by using phase imprinting technique \cite{Science.287.97}.

When there exists an abrupt variation in the potential on the way of a moving condensate, significant quantum reflection occurs, for example in the vicinity of a solid surface. Quantum reflection of cold atoms by some various structures, such as, thin films, graphene and semiconductor heterostructures \cite{NJP.13.083020} have triggered much interest both for the essential understanding of the implications of quantum mechanics and for potential applications in, e.g., the formation of zero cross-talk optical junctions \cite{PRE.53.4137}, atom chips \cite{PRL.84.4749, Nature.413.498} and a new generation of devices for precision measurement \cite{NJP.15.083002}. Matter-wave reflectors integrated with laser devices also play an important role in developing atom optics.

Most of theoretical and experimental work concern in the reflection of condensates from a solid surface at normal incidence and study the relation between the reflection probability and the incident velocity \cite{physica.d.238.1299,NJP.13.083020,PRL.93.223201,PRL.97.093201,PRA.75.022902}. Influence of interatomic interaction of atomic cloud on the reflection probability and diffraction of the condensate cloud was studied in Refs. \cite{PRL.97.093201,PRA.78.053623}. It has been shown that quantum reflection can disrupt the internal structure of condensate clouds, damping the center-of-mass motion and generating vortex excitations \cite{PRA.74.053605,PRA.81.043610}. Quantum reflections of vortices in BECs incident on a solid surface and a Gaussian tunnel barrier were studied in Refs. \cite{PRA.74.043619,PRA.75.065602}, where the authors focused on the dynamical excitations induced by interatomic interaction and gave the phase diagram of the vortex stability. The dynamics of one-dimensional dark-solitons in inhomogeneous BECs when incident on a potential gradient is studied in Ref. \cite{JPB.36.2891}.

However, the underlying physics of two-dimensional (2D) dark-soliton and barrier interaction during the reflection/transmission process has not been well described. Since the V-shaped cut of a dark-soliton can be oriented flexibly in space, an immediate issue arising is whether the reflection rate of a matter-wave soliton from a typical optical barrier depends on this orientation. In this paper, we investigate how condensates with a dark soliton are scattered off a potential wall, which allows both quantum reflection and transmission of the condensate. Our numerical simulations show that for a barrier height comparable to the initial potential of the BEC cloud, the temporal reflection rate of the matter wave is sensitive for the width of the barrier as well as the orientation of the soliton. The underlying mechanism of the quantum reflection is discussed intensively.

%
\begin{figure}[t]
  \centering
  \includegraphics[scale=0.4]{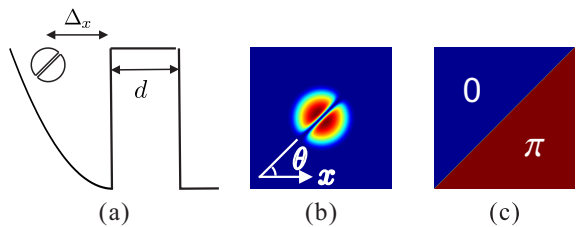}\\
  \caption{(a) The schematic view of a condensate containing a dark-soliton, whose center-of-mass is away from the rectangular potential barrier with distance $\Delta_x$ in the $x-$axis. (b) The top view of the initial density distribution of the condensate under one set of specific parameters $(\theta=\pi/4)$. (c) The corresponding phase diagram of the condensate in (b).}\label{fig1}
\end{figure}

\section{Theoretical Model}\label{sec2}
We consider a BEC containing $^{87}Rb$ atoms with mass $m$ initially confined by a harmonic trapping potential, $V_h(x,y,z)=m(\omega_x^2x^2+\omega_y^2y^2+\omega_z^2z^2)/2$, with $\omega_i$ being the trap frequency in $i=x,y,z$ direction. Here we chose the trap frequencies to be $\omega_x=\omega_y=\omega=2\pi\times5 Hz$ and $\omega_z=2\pi\times100 Hz$, where the trap frequencies in the $x$ and $y$ directions are much less than that in the $z$ direction, so the system can be regarded as a disk-shaped 2D condensate. The dynamics of the system at zero temperature is governed by the 2D time-dependent GP equation,
\begin{equation}
i\hbar\partial_t\psi=-\frac{\hbar^2}{2m}\nabla^2\psi+V_{tr}(x,y)\psi+g_{2D}N|\psi|^2\psi
\end{equation}
where $\psi(x,y,t)$ is the order parameter of the system, and $g_{2D}=2\sqrt{2\pi}\hbar\omega_za_sa_z$ is the 2D coupling constant. The s-wave scattering length $a_s=5.4nm$ and the oscillation length in the radial and axial directions are $a_0=\sqrt{\hbar/m\omega}$ and $a_z=\sqrt{\hbar/m\omega_z}$, respectively. The number of atoms, $N$, is chosen to be $10^4$ in our calculation. To prepare a condensate initially containing a dark-soliton, one can solve the GP equation in imaginary time by adding compulsively a $\pi$ phase step which divides the condensate wave function into two halves. 
We create a potential barrier which makes the total trapping potential to be in the form
\begin{equation}
  V_{tr}(x,y)=\left\{ \begin{array} {ccc} V_h(x,y,0) & x<0 \\ h &~~~~
 x\in[0, d] \\  0 & x>d \end{array} \right.
\end{equation}
with $h$ being the strength of the potential barrier and $d$ being the width of the barrier. In the middle of the harmonic trap along the $y$-axis, we place the center-of-mass of the condensate away from the trap center at $(\Delta_x, \Delta_y)$, and then let it evolve freely to study the reflection process as shown in Fig. \ref{fig1}(a). The orientation angle $\theta$ between the dark-soliton and the positive $x$-axis can be adjusted freely when the soliton is initially imprinted in the condensate. The density and phase diagram of the initial state of the condensate containing a dark-soliton with $\theta=\pi/4$ are shown in Figs. \ref{fig1}(b) and \ref{fig1}(c). In the following we use dimensionless quantities by denoting $a_0$ as the unit of length and $1/\omega$ as the unit of time.

A dark-soliton in a 2D BEC cloud is generally dynamically unstable, and it will decay eventually into vortices via snake instability  \cite{PRL.76.2262,PR.298.81}. However, we find that the lifetime of a 2D dark soliton depends strongly on the trap frequencies. The basic argument is that, for this study, the lifetime of the soliton should be larger than the timescale of the scattering process. For the given parameters in this paper, in the absence of the barrier potential, a dark-soliton sitting in the middle of the condensate cloud will not curve until at about $t=1.89T$, where $T=2\pi/\omega=200$ ms is the period of dipole oscillation of the condensate in the trap. The timescale of our calculations is about $0.25T$, which is small enough to probe how the soliton in a condensate will affect the quantum reflection. We find that the dipole oscillation of the BEC in the trap has no influence on the lifetime of the soliton. With the same parameters but a dark soliton being located at the side part of the condensate cloud, the soliton will bent and decay gradually during its oscillation from its original position. The obvious decay occurs at about $0.4T$ which is much less that of the central soltion. We also find that the quantum noise can destabilize solitons, which makes a central soliton begin to decay at about $0.39T$. These are all lager than the timescale we considered. We note that the stability of a 2D dark soliton in trapped BECs is an interesting topic on its own.

\begin{figure}[t]
  \centering
  \includegraphics[scale=0.55,clip]{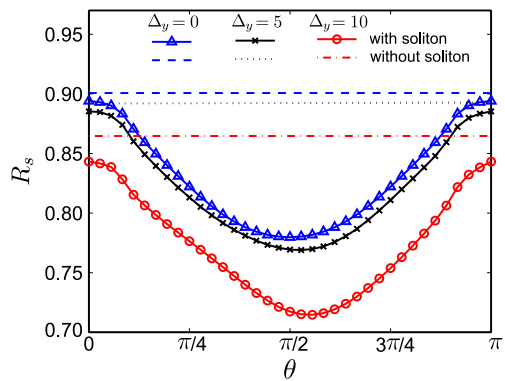}\\
  \caption{The asymptotic value of reflection probability, $R_s$, as a function of the orientation angle of soliton , $\theta$, in interacting BECs. The initial displacement of the atomic cloud in $y$ direction is labeled by different colors: $\Delta_y=0$(blue), 5(black),~10(red), while the initial displacement in $x$ direction keeps to be $\Delta_x=10$ for all lines. The solid lines with markers show the results of the systems with soliton. The value of $R_s$ for the systems in absence of soliton are given by the lines without markers for reference.}\label{fig2}
\end{figure}

\section{Results and Discussions}\label{sec3}

As stated in Refs. \cite{PRA.87.023603, LP.24.115502}, the initial displacement $(\Delta_x, \Delta_y)$ of the condensates from the trap centre plays an important role in the interference process and formation of dynamical excitations. This also applies when we consider the interaction between condensate cloud and the barrier potential. The condensate cloud will move toward the trap center, reach the potential barrier with a mean incident speed, $\bar{v} \approx\sqrt{\Delta_x^2+\Delta_y^2}$, and is then scattered by the potential barrier. At low incident velocity and high-density situation, the nonlinear interatomic interaction dominates the system, which makes the reflection process more complicated. Soliton and vortex excitations emerge more easily during the condensate cloud scattered by the potential, resulting in low reflection rate, which has been well studied. Here we address specially the influence of the width and height of the barrier, the incident angle of the condensate, and the angle between the soliton and the $x$-axis on the reflection probability and dynamical process of the system.

\begin{figure}
  \centering
  \includegraphics[scale=0.52]{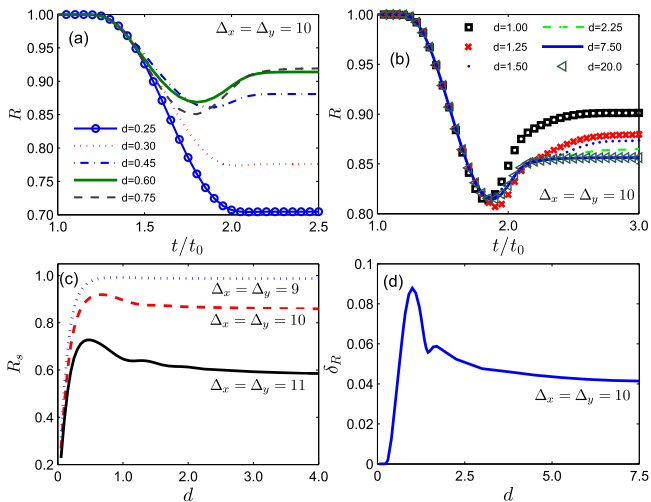}\\
  \caption{(a)(b) The reflection probabilities, $R$, as a function of time with respect to varying width of the barrier, $d$. (c) The curve of $R$ as a function of width of $d$. (d)  The difference between the minimum and the final value of $R$ as a function of $d$. In these figures, the orientation angle of the soliton is chosen to be $\theta=0$.}\label{fix_w}
\end{figure}

We introduce the reflection rate
\begin{equation}
R(t)=\frac{1}{N}\int_{-L/2}^0\int_{-L/2}^{L/2}|\psi(x,y,t)|^2dxdy
\end{equation}
with $L$ being the size of the system chosen for the calculation. When the height of the barrier potential $h$ is small, no reflection occurs. While h is sufficiently large (corresponding to a hard-wall potential), elastic reflection occurs with $R_s=1$. Without loss of generality, we chose $h=60$ and $\Delta_x=\Delta_y=10$, corresponding to the initial potential energy of $100$. 
The incident angle can be easily adjusted by varying the value of $\Delta_y$. We find that for a non-interacting system, the contribution of reflection (transmission) is mainly from the kinetic energy in the $x$-direction. We denote by $R_s$ the asymptotic reflection probability after the condensate cloud has been completely reflected by the potential barrier. There is nearly no difference of $R_s$ among various values of $\Delta_y$ for fixed $\Delta_x$ in a non-interacting system. However, when the interatomic interaction is involved, the results change greatly as shown in Fig. \ref{fig2}. One can see that $R_s$ reduces gradually with increasing $\Delta_y$ for both systems with and without soliton. In contrast to bare BEC, the BEC with a dark-soliton has a higher energy, resulting in a lower reflection rate.

We note that the asymptotic reflection probability of dark-soliton is highly sensitive to the orientation angle $\theta$. The discrepancy between $\theta = 0$ and $\theta = \pi/2$ appears to be nearly 15$\%$, which is macroscopically large for current experiments to measure. Up to now the experimentally realized reflection rate for cold atoms scattering from solid surface is nearly 70$\%$ \cite{PRL.93.223201,PRL.97.093201}. We argue that the quantum reflection rate of dark-soliton can be a signature for identifying the orientation angle of solitary waves. When the incident direction of the condensate is perpendicular to the potential surface ($\Delta_y=0$), the lowest $R_s$ appears at $\theta=\pi/2$, while for other incident angles, it shifts a little bit but still around $\theta=\pi/2$. We also find that $R_s$ for bared BEC is always higher than that of the condensate containing a soliton for a specific initial displacement, which means the soliton structure in a condensate depresses the reflection from the potential barrier.

We further study the effect of the width of the barrier, $d$, on the quantum reflection of a condensate with a dark-soliton. Figures \ref{fix_w}(a) and \ref{fix_w}(b) show that for various $d$ and fixed initial displacement $\Delta_x=\Delta_y=10$, $R(t)$ displays different features. Before the condensate contacts with the barrier potential, $R$ keeps to be 1, and then decreases monotonically during the scattering process 
until reaches its lowest value. The decrease of $R(t)$ is attributed to the transmission of the part of condensate through the barrier potential. 
However, the following variation of $R(t)$ is not the same for different values of $d$. For a narrow potential barrier (e.g., $d=0.25$, blue line with circles in Fig. \ref{fix_w}(a)), after the condensate cloud is completely separated from the potential barrier, $R$ approaches its asymptotic value, which is nearly identical to the minimum value. By contrast, for wider barrier, $R$ increases in a visible degree from the minimum up to an asymptotic value. As a result, a dip configuration in the $R-t$ curves is manifested.

To quantify the influence of the barrier width on the depth of the dip, we define $\delta_R=R_s-R_{m}$ with $R_m$ being the lowest reflection probability, which characterizes the depth of the dip of the $R-t$ curves. As shown in Fig. \ref{fix_w}(d), for $d<0.25$, $\delta_R$ is nearly zero and then increases rapidly until reaching the peak at $d=1$. For larger $d$, $R$ decreases in a damped oscillation way towards an asymptotic value. After $d>7.5$, the $R-t$ curves for different $d$ overlap as shown in Fig. \ref{fix_w}(b). In Fig. \ref{fix_w}(c), we give three $R_s-d$ curves for $\Delta_x=10=\Delta_y=9,~10,~11$, i.e., keeping the incident angle unchanged. The results show that the larger the initial displacement is, the more obvious the damped oscillation decay after $R$ increases to the maximum is. For non-interacting systems, the size of the condensate is smaller, and the dip of the $R(t)$ curve is narrower and deeper than  those of the interacting systems.

Motivated by the non-interacting quantum reflection where the reflection rate of a wave is closely related to its incident speed, we define the average momentum in the $x$-direction as $\bar{k}_x = \frac{1}{N}\int \rho(k_x, k_y) k_x dk_x$. After the condensate cloud is released, the potential energy turns into kinetic energy. The momentum of the atoms increases until they contacts with the barrier. In the subsequent process the scattered particles with opposite momentum are involved, which makes $\bar{k}_x$ reduce gradually. Through our calculations, we find that the minimum $R$ as shown in Fig. \ref{fix_w}(a) corresponds approximately to $\bar{k}_x = 0$, revealing the balance between the reflection and transmission of the matter wave. When the quantum reflection completed, $\bar{k}_x$ reaches its minimum, resulting in the plateau of the $R(t)$ curves as shown in Fig. \ref{fix_w}(a).

\begin{figure}[t]
  \centering
  \includegraphics[scale=0.6, clip]{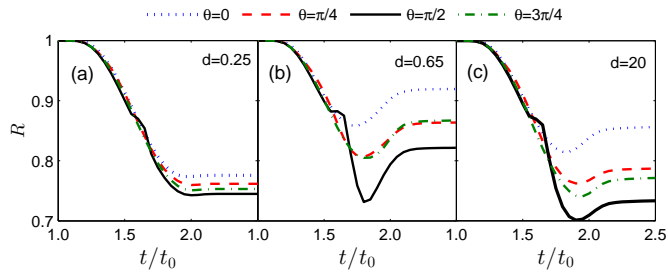}\\
  \caption{Time evolution of $R$ with respect to different soliton orientation $\theta$ for the varying width of the barrier potential, i.e., $d=0.25$ (a), 0.65 (b), and 20 (c). Blue dotted line, red dashed line, green dash-dotted line and black solid line correspond to $\theta=0, \pi/4, \pi/2$ and $3\pi/4$, respectively. In all these figures, we choose $\Delta_x=\Delta_y=10$.}\label{theta-d}
\end{figure}

The sensitivity of $R_s$ on the orientation angle of the dark-soliton has been discussed in Fig. \ref{fig2}. Now we illustrate the $R-t$ curve with some typical value of $\theta$ and explore the influence of $d$ on such sensitivity with insight. In Fig. \ref{theta-d}, we can see clearly that during the decrease of $R$ for $\theta = \pi/2$, the gradient changes suddenly due to the fact that the dark-soliton with $\theta=\pi/2$ is parallel to the potential barrier. When the soliton-barrier collision occurs, less atoms transmit through the barrier.
For any value of $d$, $R_s$ for $\theta=0$ is always the largest, while $R_s$ for $\theta=\pi/2$ is always the smallest. Due to the symmetry of the systems with $\theta=\pi/4$ and $3\pi/4$, $R_s$ of these two configurations can be nearly the same as shown in Fig. \ref{theta-d}(b). For other values of $d$ when $0\leq\theta<\pi$, the smaller $\theta$ is, the greater $R_s$ is, as shown in Figs. \ref{theta-d}(a) and \ref{theta-d}(c). We can also see that with increasing $d$, the distinction of $R_s$ for $\theta = 0$ and $\theta = \pi/2$ increases significantly. The value of $R_s(\theta = 0) - R_s(\theta = \pi/2)$ is 0.031 for $d = 0.25$ and 0.1225 for $d = 20$. It indicates that the width of the barrier tends to enhance the quantum reflection sensibility. Our results suggest experiments to identify precisely the direction of the dark-soliton in condensates and explore its dynamics through the quantum reflection of BEC on the barrier potential  since the distinction of $R_s$ due to the incident direction of the matter wave soliton can be up to more than 10 percent of total particles and measured easily in current experiments.

\begin{figure}[t]
  \centering
  \includegraphics[scale=0.7, clip]{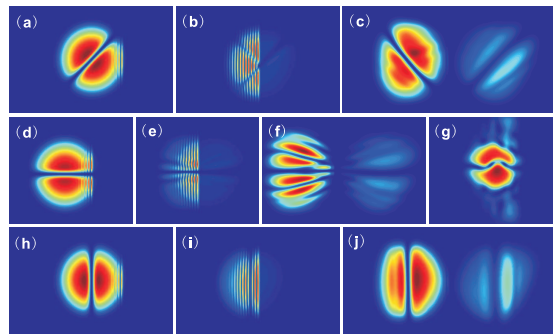}\\
  \caption{Density profiles of condensate with the soliton orientation angle $\theta=\pi/4$ (a)-(c), $\theta=0$ (d)-(g) and $\theta=\pi/2$ (h)-(j), during the reflection process. The barrier width is $d=0.25$, and the initial displacement is $\Delta_x=\Delta_y=10$. Time for plots (a), (d) and (h) is $t=0.19T$. Times for plots (b), (e) and (i) are $t=0.26T, 0.27T$ and $0.23T$, respectively. Time for plots (c), (f) and (j) is $t=0.34T$. Time for plot (g) is $t=0.63T$. }\label{dens-evol}
\end{figure}

The reflection process also provides a platform to study the interaction between dark-soliton and barrier potential and related nonlinear effects. In Fig. \ref{dens-evol}, we show the temporal density distribution for $\theta =\pi/4$ (the first row), $\theta = 0$ (the second row) and $\theta = \pi /2$ (the third row). After interacting with the barrier potential, the incident cloud splits into two components, reflected and transmitted waves. We emphasize that both counter parts contain a dark-soliton structure, but they do not fulfill mirror symmetry except for $\theta = 0$ (see Figs. \ref{dens-evol}(c), \ref{dens-evol}(f) and \ref{dens-evol}(j)). For the cases of $\theta = \pi/4$ and $\pi/2$, the dark-soliton structures appear stably in the reflected clouds after the scattering process (see Figs. \ref{dens-evol}(c) and \ref{dens-evol}(j)). However, we note that the position of these solitons turn to be off-centre. For $\theta=0$, interference along $y$-axis is obvious as shown in Fig. \ref{dens-evol}(f) after collision, and the initial central soliton breaks and evolves into vortex pairs via snake instability quickly, which makes the reflected cloud highly excited (see Fig. \ref{dens-evol}(g)). This indicates that the head-on collision between the incident and reflected parts of the soliton induces strong sound wave excitations and then breaks the dark-soliton structure.

We also find that the wider the barrier is, the stronger the excitations are in the condensate cloud when the reflection is finished. Moreover, for large $d$ with
any initial orientation angle, the soliton tends to decay into vortices as shown in Fig. \ref{dens-evol}(g), and the dark-soliton structure do not appear in the transmitted cloud anymore.

\section{Conclusion}\label{sec4}
The quantum reflection of BECs containing a dark-soliton from a barrier potential is affected by several factors, including the barrier width and height, the initial displacement of the condensate cloud, and especially the orientation angle of the dark-soliton with respect to the barrier surface. As the perfect initial dark-soliton structure (without some residual sound excitations) obtained by the imaginary time evolution may not be able to produce in experiments, we suggest to prepare these 2D configurations by combining condensates in a double well potential with phase imprinting technique as described in Ref. \cite{PRA.87.023603}, where the created dark-soliton can also survive for $1.19T$. We find that the orientation of the soliton affects the quantum reflection in a non-trivial way and the sensitivity of the reflection probability on the orientation of the soliton is significant. This sensitivity may be used in probing surfaces or the internal structure of dark-solitons. Although the $\theta=0$ dark-soliton is less stable than the dark-soliton with other orientations after the reflection, the asymptotic reflection probability of the system is always the largest. However, for the same orientation, the interatomic interaction and the soliton structure enhance the transmission. The larger the velocity in the $y$-direction induced by the larger $\Delta_y$ is, the more contribution to the motion of the condensate in the $x$-direction by the nonlinear interaction, i.e., the nonlinear effects transfer the momentum in the $y$-direction to the $x$-direction and enhance the transmission of the cloud. These results enrich our understanding of BEC-surface interactions and contribute to building new atom devices.

\section*{Acknowledgments}

We thank German Sinuco for useful discussions. Supports from National Natural Science Foundation of China (11775178 and 11775177), Major Basic Research Program of Natural Science of Shaanxi Province (2017KCT-12, 2017ZDJC-32) and The Double First-class University Construction Project of Northwest University are acknowledged.


\end{document}